\documentclass{ws-ijmpcs}

\begin{document}

\markboth{Bo E. Sernelius}
{Casimir Effects in Graphene Systems}

\catchline{}{}{}{}{}
%
%%%%%%%%%%%%%%%%%%%%%%%%%%%%%%%%%%%%%%%%%%%%%%%%%%%%%%%%%%%%%%%%%%%%

\title{CASIMIR EFFECTS IN GRAPHENE SYSTEMS: UNEXPECTED POWER LAWS}

\author{BO E. SERNELIUS}

\address{Department of Physics, Chemistry and Biology,
Link\"{o}ping University, \\SE-581 83 Link\"{o}ping, Sweden\\
bos@ifm.liu.se}

\maketitle

\begin{history}
\received{Day Month Year}
\revised{Day Month Year}
\end{history}

\begin{abstract}
We present calculations of the zero-temperature Casimir interaction between
two freestanding graphene sheets as well as between a graphene sheet and a substrate.
Results are given for undoped graphene and for a set of doping levels covering the range of experimentally accessible values. We describe different approaches that can be used to derive the interaction. We point out both the predicted power law for the interaction and the actual distance dependence.

\end{abstract}

\ccode{PACS numbers: 73.21.-b, 73.22.Pr}

\section{Introduction}	
Graphene\,\cite{Boehm}\cdash\cite{Novo2} has a very special band structure, leading to unexpected results. G$\acute{o}$mez-Santos\,\cite{Santos} showed that retardation effects are negligible for un-doped graphene. This is one of the peculiarities caused by the band structure. This means that we can limit the calculations to van der Waals interactions which are easier to derive. Throughout the text we make the basic derivations more general so that they are valid for any type of two-dimensional (2D) sheets, like, e.g., 2D metallic sheets. The result becomes specific for graphene when we insert the graphene polarizability, $\alpha  \left( {q,\omega } \right) $. In terms of the polarizability the dielectric function is given by
$
\varepsilon \left( {q,\omega } \right) = 1 + \alpha  \left( {q,\omega } \right) = 1 - v^{2D} \left( q \right){{\chi \left( {q,\omega } \right)}}$, where $v^{2D} \left( q \right) = {{2\pi e^2 } \mathord{\left/ {\vphantom {{2\pi e^2 } q}} \right. \kern-\nulldelimiterspace} q}$ is the 2D Fourier transform of the coulomb potential and ${\chi \left( {q,\omega } \right)}$ the density-density correlation function or polarization bubble. For undoped graphene, in a general point in the complex frequency plane, away from the real axis the density-density correlation function is\cite{Guinea}
%1
\begin{equation}
\chi \left( {{\bf{q}},z} \right) =  - \frac{g}{{16\hbar }}\frac{{q^2 }}{{\sqrt {v^2 q^2  - z^2 } }},
\label{equ1}
\end{equation}
where $v$ is the carrier velocity which is a constant in graphene ($E =  \pm \hbar vk$), and $g$ represents the degeneracy parameter with the value of four (a factor of two for spin and a factor of two for the cone degeneracy.) For doped graphene the function becomes much more complicated\,\cite{Wun}\cdash\cite{SerEpl}. We refer the reader to Ref.\,\refcite{SerEpl} where the expression on the imaginary axis can be found in terms of real valued functions of real valued variables. Other calculations where graphene were modeled differently has been publihed\,\cite{Bordag2,Woods}.

Section 2 is devoted to two freestanding graphene sheets, undoped or doped. We present derivations of the interaction and numerical results. In section 3 we treat a graphene sheet above a substrate. We derive the interaction and present results for a gold substrate. In section 4 we discuss the predicted power law dependence of the interactions and compare to the actual outcome from our calculations. Finally, section 5 we devote to summary and conclusions. Part of his work has previously been reported in a letter\,\cite{SerEpl}.
\begin{figure}[pb]
\centerline{\psfig{file=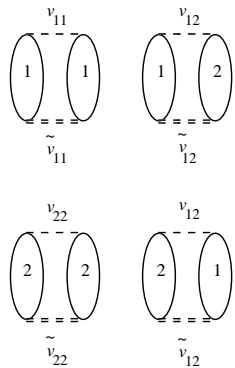,width=3cm}}
\vspace*{8pt}
\caption{Feynman diagrams for the correlation energy in the two graphene sheet system. The ellipses represent polarization bubbles and the dashed lines the interactions indicated in the figure. The numbers 1 and 2 refer to which sheet the electron belongs to. See Ref. 15 for details.}
\label{figu1}
\end{figure}
\section{Two Freestanding Graphene Sheets}
It is possible\,\cite{Ser1}\cdash\cite{Svet1} but very difficult to take spatial dispersion into account when calculating Casimir interactions in systems with three-dimensional (3D) layers. When dealing with parallel 2D sheets, like in the case of two freestanding graphene sheets,  spatial dispersion enters in a natural way and does not cause any complications. Thus these systems can be used to gain experience concerning possible effects of spatial dispersion in general.
\subsection{Derivations}
The derivation of the non-retarded Casimir interaction can be done in several different ways. Here, we will present two different derivations; one   in the language of many-body theory; one in terms of electromagnetic normal modes. We begin with the many-body treatment.
\subsubsection{Many-body theory}

For a system of parallel 2D sheets the derivation in the language of many-body theory becomes very simple, especially when retardation effects can be neglected. In the present two-sheet system the interaction energy is nothing but the inter-sheet correlation energy\,\cite{SerBjo}. Using diagrammatic perturbation theory the Feynman diagrams representing the correlation energy are given in Fig.\,\ref{figu1}. To get the inter-sheet contribution we can either subtract the intra-sheet part or subtract the result when the separation between the sheets goes to infinity (at that limit only the intra-sheet contribution remains.) To avoid unnecessary complications we assume that the two sheets are identical. If we have two doped graphene layers the doping concentration is the same in the two sheets. It is straight forward to extend the treatment to different sheets.
Each ellipse represents a polarization bubble,  $\chi \left( {{\bf{q}},\omega } \right)$, and the number indicates in which sheet the process occurs. Each of the four diagrams represent an infinite series of diagrams. Since the sheets are identical the diagrams in the second row give  contributions identical to the diagrams in the first row. We can use the first row diagrams and multiply the result with a factor of two. The interaction energy per unit area can be written as
%2
\begin{equation}
{E_c}\left( d \right) = \hbar \int {\frac{{{d^2}q}}{{{{\left( {2\pi } \right)}^2}}}} \int\limits_0^\infty  {\frac{{d\omega }}{{2\pi }}\int\limits_0^1 {d\lambda \frac{1}{\lambda }} 2\left[ {Diag1\left( {q,i\omega ;\lambda } \right) + Diag2\left( {q,i\omega ;\lambda } \right)} \right]}, 
\label{equ2}
\end{equation}
where $\lambda$ is the coupling constant and the factor of two has been inserted. 

Now ${v_{11}} = {v^{2D}}\left( {{\bf{q}},\omega } \right)$ and ${v_{12}} = \exp \left( { - qd} \right){v^{2D}}\left( {{\bf{q}},\omega } \right)$, respectively, where $d$ is the distance between the sheets. The exponential factor in the second potential is the result from taking the 2D Fourier transform of the coulomb potential in a plane the distance $d$ from the center of the potential. The interaction lines with double bars represent a series of terms, with zero, one, two $ \ldots $ number of polarization bubbles. This can be expressed as
%3
\begin{eqnarray}
\nonumber {{\tilde v}_{11}}\left( {{\bf{q}},\omega } \right) &=& {v^{2D}}\left( q  \right) + {v^{2D}}\left( q  \right)\chi \left( {{\bf{q}},\omega } \right){{\tilde v}_{11}}\left( {{\bf{q}},\omega } \right) + \exp \left( { - qd} \right){v^{2D}}\left( q  \right)\chi \left( {{\bf{q}},\omega } \right){{\tilde v}_{12}}\left( {{\bf{q}},\omega } \right)\\
\nonumber {{\tilde v}_{12}}\left( {{\bf{q}},\omega } \right) &= &\exp \left( { - qd} \right){v^{2D}}\left( q  \right) + {v^{2D}}\left( q  \right)\chi \left( {{\bf{q}},\omega } \right){{\tilde v}_{12}}\left( {{\bf{q}},\omega } \right)\\
& &+ \exp \left( { - qd} \right){v^{2D}}\left( q  \right)\chi \left( {{\bf{q}},\omega } \right){{\tilde v}_{11}}\left( {{\bf{q}},\omega } \right),
\label{equ3}
\end{eqnarray}
where we have closed the two infinite series. This system of equations can be solved and the result is
%
%4
\begin{eqnarray}
\nonumber {{\tilde v}_{11}}\left( {{\bf{q}},\omega } \right) = \frac{{{v^{2D}}\left( q \right)\left[ {1 + \alpha \left( {{\bf{q}},\omega } \right)\left( {1 - \exp \left( { - 2qd} \right)} \right)} \right]}}{{{{\left[ {1 + \alpha \left( {{\bf{q}},\omega } \right)} \right]}^2} - \exp \left( { - 2qd} \right){\alpha ^2}\left( {{\bf{q}},\omega } \right)}}\\
{{\tilde v}_{12}}\left( {{\bf{q}},\omega } \right) = \frac{{{v^{2D}}\left( q \right)\exp \left( { - qd} \right)}}{{{{\left[ {1 + \alpha \left( {{\bf{q}},\omega } \right)} \right]}^2} - \exp \left( { - 2qd} \right){\alpha ^2}\left( {{\bf{q}},\omega } \right)}}.
\label{equ4}
\end{eqnarray}
The square brackets in Eq.\,(\ref{equ2}) are $\left[ {{{\tilde v}_{11}}{v_{11}}{\chi ^2} + {{\tilde v}_{12}}{v_{12}}{\chi ^2} - \left( {{v^{2D}}/\varepsilon } \right){v^{2D}}{\chi ^2}} \right]$, where the last term comes from the subtraction of the intra-band correlation energy. Each factor of ${e^2}$ appearing implicitly in the expression should be multiplied by the coupling constant. Performing the integration over coupling constant gives
%5
\begin{equation}
{E_c}\left( d \right) = \hbar \int {\frac{{{d^2}q}}{{{{\left( {2\pi } \right)}^2}}}\int\limits_0^\infty  {\frac{{d\omega }}{{2\pi }}\ln \left\{ {1 - {e^{ - 2qd}}{{\left[ {\frac{{\alpha \left( {q,i\omega } \right)}}{{1 + \alpha \left( {q,i\omega } \right)}}} \right]}^2}} \right\}} }.
\label{equ5}
\end{equation}
This is the result from diagrammatic perturbation theory within the random phase approximation (RPA). In next section we derive the same thing using normal modes.
\subsubsection{Normal-mode derivation}
An electromagnetic normal mode\,\cite{Ser3} is a solution to Maxwell's equations in absence of external perturbations. At zero temperature the interaction energy of the system is the change in the total zero-point energy of all the normal modes when interaction is turned on. In a planar system like the ones we treat in this work the modes are characterized by the 2D wave vector $\bf q$ in the plane of the sheets and substrates. Within this formalism the interaction energy is obtained as
%6
\begin{equation}
{E_c}\left( d \right) = \hbar \int {\frac{{{d^2}q}}{{{{\left( {2\pi } \right)}^2}}}\int\limits_0^\infty  {\frac{{d\omega }}{{2\pi }}\ln \left\{ {{f_{\bf{q}}}\left( {i\omega } \right)} \right\}} },
\label{equ6}
\end{equation}
where ${f_{\bf{q}}}\left( {{\omega _{\bf{q}}}} \right) = 0$ is the condition for electromagnetic normal modes. One arrives at this expression after using an extension of the so-called argument principle and deforming the integration path in the complex frequency plane.\,\cite{Ser3}

We will now go through one possible way to find the normal modes and their zero-point energies. Let us assume that we have an induced carrier distribution, $\rho_1 \left( {{\bf{q}},\omega } \right)$, in sheet 1. This gives rise to the potential $v\left( {{\bf{q}},\omega } \right) = {{v^{2D} \left( q \right)\rho _1 \left( {{\bf{q}},\omega } \right)} }$ and ${{\exp \left( { - qd} \right)v^{2D} \left( q \right)\rho _1 \left( {{\bf{q}},\omega } \right)}}$
 in sheets 1 and 2, respectively. The resulting potential in sheet 2 after screening by the carriers is $\exp \left( { - qd} \right){v^{2D}}\left( q \right){\rho _1}\left( {{\bf{q}},\omega } \right)/\left[ {1 + \alpha \left( {{\bf{q}},\omega } \right)} \right]$, which gives rise to an induced carrier distribution in sheet 2, 
%7
\begin{equation}
\rho _2 \left( {{\bf{q}},\omega } \right) = \chi \left( {{\bf{q}},\omega } \right)e^{ - qd} v^{2D} \left( q \right)\frac{{\rho _1 \left( {{\bf{q}},\omega } \right)}}{{\ \left[ {1 + \alpha  \left( {{\bf{q}},\omega } \right)} \right]}}.
\label{equ7}
\end{equation}
In complete analogy, this carrier distribution in sheet 2 gives rise to a carrier distribution in sheet 1,
%8
\begin{equation}
\rho _1 \left( {{\bf{q}},\omega } \right) = \chi \left( {{\bf{q}},\omega } \right)e^{ - qd} v^{2D} \left( q \right)\frac{{\rho _2 \left( {{\bf{q}},\omega } \right)}}{{ \left[ {1 + \alpha  \left( {{\bf{q}},\omega } \right)} \right]}}.
\label{equ8}
\end{equation}
\begin{figure}[pb]
\centerline{\psfig{file=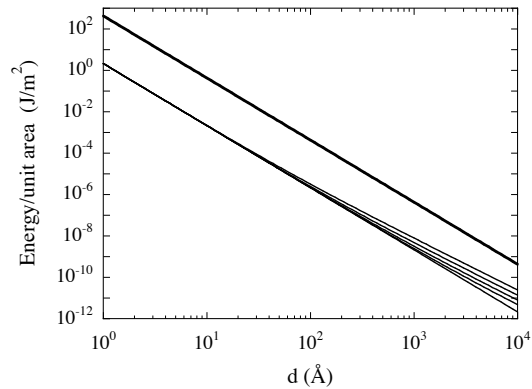,width=7cm}}
\vspace*{8pt}
\caption{The attractive interaction energy between two graphene sheets. The lower straight line is for undoped sheets, while the four bent curves are for doping densities $1 \times 10^{10}, {\rm{ }}1 \times 10^{11}, {\rm{ }}1 \times 10^{12}, {\rm{ and }} {\kern 1pt} 1 \times 10^{13} {\,\rm{ cm }}^{-2} $, respectively, counted from below. The uppermost thick straight line is the classical Casimir result for two perfectly reflecting half spaces. \label{figu2}}
\end{figure}
To find the condition for self-sustained fields, normal modes, we let this induced carrier density in sheet 1 be the carrier density we started from. This leads to
%9
\begin{equation}
1 - e^{ - 2qd} \left[ {\frac{{\alpha  \left( {{\bf{q}},\omega } \right)}}{{1 + \alpha  \left( {{\bf{q}},\omega } \right)}}} \right]^2  = 0.
\label{equ9}
\end{equation}
The left hand side of this equation is exactly the argument of the logarithm in Eq.\,(\ref{equ5}). That equation was derived using many-body theory. 

\subsection{Results}
The numerical results for the size of the interaction energy between two undoped graphene layers in vacuum is shown as the lower straight line in Fig.\,\ref{figu2}. The bent curves are valid for doping concentrations  $1 \times 10^{10},{\rm{ }}1 \times 10^{11},{\rm{ }}1 \times 10^{12}, {\rm{ and }}{\kern 1pt}  1 \times 10^{13} { \,\rm{ cm }}^{-2} $, respectively, counted from below. The interaction energy is negative, leading to an attractive force. The uppermost thick straight line is the classical Casimir result for two perfectly reflecting half spaces. We note that the two straight lines are parallel and will never cross. The curves representing doped graphene sheets on the other hand will, for distances well outside the region covered here, approach and join the classical Casimir result.

\section{A Graphene Sheet Above a Substrate}
\begin{figure}[pb]
\centerline{\psfig{file=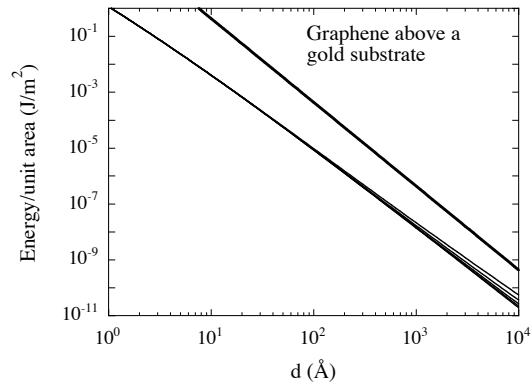,width=7cm}}
\vspace*{8pt}
\caption{The attractive interaction energy between a graphene sheet and a gold substrate. The lowest curve is for an undoped sheet, while the the following four curves are for doping densities $1 \times 10^{10}, {\rm{ }}1 \times 10^{11}, {\rm{ }}1 \times 10^{12}, {\rm{ and }} {\kern 1pt} 1 \times 10^{13} {\,\rm{ cm }}^{-2} $, respectively, counted from below. The uppermost thick straight line is the classical Casimir result for two perfectly reflecting half spaces. \label{figu3}}
\end{figure}

\subsection{Derivations}
This geometry is not well suited for using many-body theory so we go straight to an approach using electromagnetic normal modes.
\subsubsection{Normal-mode derivation}
If we have a 2D layer (like a graphene sheet) the distance $d$ above a substrate the procedure is very similar to in Sec. (2.1.2). We start with an induced mirror carrier density,  $\rho_1 \left( {{\bf{q}},\omega } \right)$, in the substrate. The induced carrier density in the 2D layer is given by the expression in Eq.\,(\ref{equ7}) except that now the distance between the mirror charge and the 2D layer is $2d$ instead of $d$. Eq.\,(\ref{equ8}) is then replaced by 

%10
\begin{equation}
\rho _1 \left( {{\bf{q}},\omega } \right) =  - \rho _2 \left( {{\bf{q}},\omega } \right)\frac{{\varepsilon _s \left( \omega  \right) - 1}}{{\varepsilon _s \left( \omega  \right) + 1}},
\label{equ10}
\end{equation}
and the condition for normal modes becomes
%11
\begin{equation}
1 - e^{ - 2qd} \frac{{\alpha  \left( {{\bf{q}},\omega } \right)}}{{1 + \alpha  \left( {{\bf{q}},\omega } \right)}}\frac{{\varepsilon _s \left( \omega  \right) - 1}}{{\varepsilon _s \left( \omega  \right) + 1}} = 0.
\label{equ11}
\end{equation}
resulting in the energy
%

%12
\begin{equation}
\begin{array}{*{20}{l}}
{{E_c}\left( d \right) = \hbar \int {\frac{{{d^2}q}}{{{{\left( {2\pi } \right)}^2}}}} \int\limits_0^\infty  {\frac{{d\omega }}{{2\pi }}} \ln \left\{ {1 - {e^{ - 2qd}}\left[ {\frac{{\alpha '\left( {q,\omega } \right)}}{{1 + \alpha '\left( {q,\omega } \right)}}\frac{{{\varepsilon _{s'}}\left( \omega  \right) - 1}}{{{\varepsilon _{s'}}\left( \omega  \right) + 1}}} \right]} \right\}.}
\end{array}
\label{equ12}
\end{equation}
\subsection{Results}
\begin{figure}[pb]
\centerline{\psfig{file=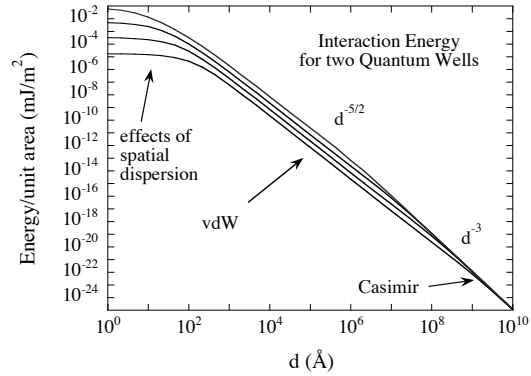,width=7cm}}
\vspace*{8pt}
\caption{The attractive interaction energy between two 2D metallic sheets. The curves are for doping densities $1 \times 10^{10}, {\rm{ }}1 \times 10^{11}, {\rm{ }}1 \times 10^{12}, {\rm{ and }} {\kern 1pt} 1 \times 10^{13} {\,\rm{ cm }}^{-2} $, respectively, counted from below. For large separations the curves follow the classical Casimir result for two perfectly reflecting half spaces; for intermediate separations, in the van der Waals region, the curves varies as $d^{-5/2}$; for separations smaller than the Thomas-Fermi screening length the interaction shows a weakening due to spatial dispersion effects See Ref. 15 for details.
\label{figu4}}
\end{figure}
\begin{figure}[pb]
\centerline{\psfig{file=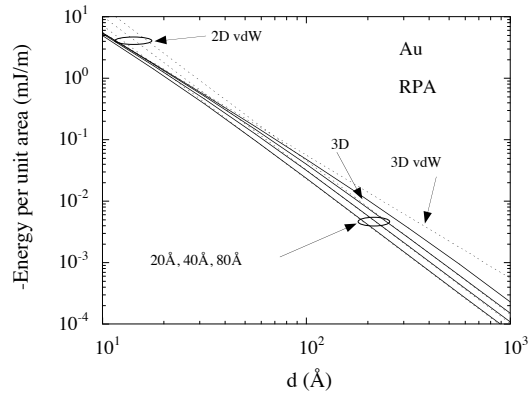,width=7cm}}
\vspace*{8pt}
\caption{The attractive interaction energy between two thin gold films. For large separations the curves follow the classical Casimir result for two perfectly reflecting half spaces; for small separations they follow the van der Waals result for two gold half spaces; in a region in between these regimes thin films behave as strictly 2D metal films in the van der Waals regime with a $d^{-5/2}$ separation dependence. See Ref. 20 for details.\label{figu5}}
\end{figure}

The result for a graphene sheet above a gold substrate is shown in Fig.\,\ref{figu3}. The dielectric function of gold along the imaginary frequency axis was obtained from experimental data extrapolated in a way described in Ref. \refcite{BosSer} and by the use of a modified Kramers Kronig dispersion relation (see Eq.\,(6.75) in Ref. \refcite{Ser3}.) 

\section{Predicted Power Laws and Actual Distance Dependence}
One way to find fast results for the van der Waals and Casimir interactions between objects of various shapes is to sum over pair interactions. According to Langbein\,\cite {Lang} one finds the correct separation dependence but the overall strength is not always right. For two parallel thin films this gives the power law $d^{ - 4}$ ($d^{ - 5}$) in the non-retarded (retarded) limit. We found in Ref. \refcite{SerBjo} that this is not true for a pair of  2D metallic sheets. We found a fractional power law, $d^{ - {5 \mathord{\left/ {\vphantom {5 2}} \right. \kern-\nulldelimiterspace} 2}}$, in the non-retarded limit (see the straight part of the solid curves in Fig.\,\ref{figu4}) and the power law $d^{ - 3}$ in the retarded. It is also interesting to note that the interaction weakens in the small separation limit for separations of the order of the Thomas-Fermi screening length. This effect is a direct result of spatial dispersion. The importance of spatial dispersion in this geometry was also found by Gerlach\,\cite{Gerl}. Later we studied 3D gold films of different thickness. We found\,\cite{BosSer2} that there is a region around the cross-over point between van der Waals interaction and Casimir interaction where thin metal films behave as strictly 2D films (See Fig.\,\ref{figu5}.) 
Here, in the present work we have a system with yet another separation dependence. For a pair of undoped graphene layers the non-retarded interaction varies as $d^{ - 3}$, verifying the prediction by Dobson et al.\,\cite{Dobson}, and the same power law holds in the retarded regime. This is the same power law as for the retarded interaction between two half spaces. The origin of the half-integer behavior for the 2D metal sheets is the square root dependence of the dispersion curve for 2D plasmons. In doped graphene the plasmon dispersion curve attains the square root dependence and for larger separations the interaction varies as $d^{ - {5 \mathord{\left/ {\vphantom {5 2}} \right. \kern-\nulldelimiterspace} 2}}$ (see the rightmost part of Fig.\,\ref{figu2}). For a graphene sheet above a metal substrate we find a more complicated behavior, i.e., no simple power law. Summation over pair interactions suggests a  $d^{ - 3}$ dependence in the non-retarded regime.

\begin{table}[ph]
\tbl{Predictions {\`a} la Langbein (within parentheses) and outcome}
{\begin{tabular}{@{}cccc@{}} \toprule
Geometry& van der Waals & Casimir \\
 \colrule
2D metal---metal half space\hphantom{00}\ & \hphantom{0}${d^{ - {5 \mathord{\left/
 {\vphantom {5 2}} \right.
 \kern-\nulldelimiterspace} 2}}}$$^{(a)}$ (${d^{ - 3}}$)& \hphantom{0} ${d^{ - 3}}$$^{(a)}$ (${d^{ - 4}}$) \\
2D metal---2D metal\hphantom{00000000} & \hphantom{0}${d^{ - {5 \mathord{\left/
 {\vphantom {5 2}} \right.
 \kern-\nulldelimiterspace} 2}}}$$^{(a,b)}$ (${d^{ - 4}}$)& \hphantom{0} ${d^{ - 3}}$$^{(a,b)}$ (${d^{ - 5}}$) \\
graphene---metal half space\hphantom{00} & \hphantom{0}No pure power law (${d^{ - 3}}$)& \hphantom{0} ${d^{ - 3}}$ (${d^{ - 4}}$) \\
graphene---graphene\hphantom{00000000} & \hphantom{0}${d^{ - 3}}$ $^{(c,d)}$(${d^{ - 4}}$)& \hphantom{0} ${d^{ - 3}}$ (${d^{ - 5}}$) \\
doped\hphantom{00000000} & \hphantom{0}${d^{ - {5 \mathord{\left/
 {\vphantom {5 2}} \right.
 \kern-\nulldelimiterspace} 2}}}$$^{(c)}$ (${d^{ - 4}}$)& \hphantom{0} ${d^{ - 3}}$ (${d^{ - 5}}$) \\
 \botrule\\

(a) Bostr\"{o}m and Sernelius, \hphantom{000} &{\it Phys. Rev. B} {\bf 61}, 2204 (2000).\hphantom{0000000}&\\
(b) Sernelius and Bj\"{o}rk,\hphantom{000000} &{\it Phys. Rev. B} {\bf 57}, 6592 (1998).\hphantom{0000000} &\\
(c) Bo E. Sernelius,\hphantom{0000000000} &{\it EPL} {\bf 95}, 57003 (2011), (present work).\hphantom{}&\\
(d) Predicted by Dobson et al., &{\it Phys. Rev. Lett.} {\bf 96}, 073201 (2006).\hphantom{000}&\\

\end{tabular} \label{ta1}}
\end{table}

Our calculations were done for $T=0$. We have so far refrained from taking finite temperatures into account since then one has to use a finite temperature dielectric function for graphene, a function that has to be calculated numerically. In Ref.\,\refcite{Fial} one recently used a quantum-field-theory model to derive the retarded interaction at finite temperature. One put the main emphasis on the high temperature limit (or the large separation limit at finite temperature.) One found that the results for two freestanding graphene sheets and for a graphene sheet above a metal substrate both agree with the result for two Drude type metal half spaces; this interaction is half the strength of the interaction between two ideal metal half spaces and varies as $d^{ - 2}$. The result agrees with that of G$\acute{o}$mez-Santos\,\cite{Santos} but there are some disagreements as to when this high-temperature result sets in. Another even more recent publication\,\cite{Sara} treats the retarded interaction between two freestanding graphene sheets, undoped and doped, at finite temperature. In the undoped case they find for small separations a $d^{ - 3}$ dependence in agreement with our results as well as with those of G$\acute{o}$mez-Santos\,\cite{Santos} and  those of Dobson {\it et al.}\,\cite{Dobson}; at large separations they find a  $d^{ - 4}$ dependence in disagreement with G$\acute{o}$mez-Santos\,\cite{Santos}. In the doped case  they find for small separations a $d^{ - 5/2}$ dependence and for large separations a  $d^{ -2}$ dependence. This is in disagreement with our results in that we find the $d^{ - 5/2}$ dependence sets in at a much larger separation. We believe the disagreements are caused by their treatment of the graphene sheets. They use the formalism for layers of finite width. They let each graphene layer keep a small but finite width and let the bulk dielectric function be represented by the 2D graphene dielectric function. We believe this is not the correct way to proceed. It produces an effective 2D dielectric function, different from the graphene dielectric function. In Ref.\,\refcite{BosSer2} we used the bulk dielectric function for gold films of finite thickness. When the film thickness goes towards zero the result approaches, as it should, that for a strictly 2D metal film. Had we instead used the dielectric function of a 2D metal film for the bulk dielectric function the result would have been quite different.

\section{Summary and Conclusions}
We have derived the non-retarded Casimir interaction between two parallel 2D sheets and performed numerical calculations for two undoped and doped graphene sheets. Spatial dispersion was fully included. Spatial dispersion has strong effect on the contribution from the doping carriers but negligible effect on the contribution from undoped graphene. We have furthermore derived the non-retarded Casimir interaction between a 2D sheet and a substrate. Numerical results are presented for an undoped and doped  graphene sheet above a gold substrate. In both geometries the power law for the separation dependence of the interaction is quite different from that predicted by Langbein. The distance dependence is summarized in Table \ref{ta1}.

\section*{Acknowledgments}
Financial support from the Swedish Research Council is gratefully acknowledged.

\end{document}